\documentclass[cernpreprint,texlive=2016,txfonts,UKenglish]{latex/atlasdoc}
\pdfoutput=1
\usepackage[biblatex=true]{latex/atlaspackage}
\usepackage{latex/atlasphysics}

\newcommand{\AtlasCoordFootnote}{
ATLAS uses a right-handed coordinate system with its origin at the
nominal interaction point (IP) in the centre of the detector and the
$z$-axis along the beam pipe. The $x$-axis points from the IP to the
centre of the LHC ring, and the $y$-axis points upwards. Cylindrical
coordinates $(r,\phi)$ are used in the transverse plane, $\phi$ being
the azimuthal angle around the $z$-axis. The pseudorapidity is defined
in terms of the polar angle $\theta$ as $\eta = -\ln
\tan(\theta/2)$. Angular distance is measured in units of $\Delta R
\equiv \sqrt{(\Delta\eta)^{2} + (\Delta\phi)^{2}}$. The transverse energy
is defined as $E_{\mathrm T}=E\sin\theta$, where $E$ is the energy.}

\graphicspath{{logos/}}

\usepackage{latex/atlasbiblatex}

\def\fb1{fb$^{-1}$}
\def\pb1{pb$^{-1}$}
\def\qq{q\bar q}

\def\etg{E_{\mathrm{T}}^{\gamma}}
\def\etag{\eta^{\gamma}}

\def\etiso{E_{\mathrm{T}}^{\mathrm{iso}}}
\def\etisocutn{$4.8+4.2\cdot 10^{-3}\cdot\etg$}

\def\sher{{\textsc{Sherpa}}}
\def\pyt{{\textsc{Pythia}}}
\def\jetp{{\textsc{Jetphox}}}

\mathchardef\mhyphen="2D

\def\figdir{figures/}

\AtlasTitle{Measurement of the cross section for inclusive isolated-photon
        production in {\boldmath $pp$} collisions
        at {\boldmath $\sqrt s=13$}~\TeV\ using the ATLAS detector}

\author{The ATLAS Collaboration}

\AtlasJournalRef{Phys. Lett. B 770 (2017) 473}
\AtlasDOI{10.1016/j.physletb.2017.04.072} 

\AtlasAbstract{
Inclusive isolated-photon production in $pp$ collisions at a
centre-of-mass energy of $13$~\TeV\ is studied with the ATLAS detector
at the LHC using a data set with an integrated luminosity of
$3.2$~\fb1. The cross section is measured as a function of the photon
transverse energy above $125$~\GeV\ in different regions of photon
pseudorapidity. Next-to-leading-order perturbative QCD and Monte Carlo
event-generator predictions are compared to the cross-section
measurements and provide an adequate description of the data.
}

\usepackage[notindex,nottoc,notlof,notlot]{tocbibind}

\begin{document}

\maketitle

\section{Introduction}
\label{intro}
The production of prompt photons in proton--proton ($pp$) collisions,
$pp\rightarrow\gamma\ +\ {\mathrm{X}}$, provides a testing ground for
perturbative QCD (pQCD) with a hard colourless probe. All photons
produced in $pp$ collisions that are not secondaries from hadron
decays are considered as ``prompt''. Two processes contribute to
prompt-photon production in $pp\rightarrow\gamma\ +\ {\mathrm{X}}$: the
direct process, in which the photon originates directly from the hard
interaction, and the fragmentation process, in which the photon is
emitted in the fragmentation of a high transverse momentum 
($p_{\mathrm{T}}$) parton~\cite{pr:d76:034003,pr:d79:114024}. Measurements
of inclusive prompt-photon production were used recently to
investigate novel approaches to the description of parton
radiation~\cite{pr:d94:034020} and the importance of resummation of
threshold logarithms in QCD and of the electroweak
corrections~\cite{jhep:1609:005}. Comparisons of prompt-photon data
and pQCD are usually limited by the theoretical uncertainties
associated with the missing higher-order terms in the perturbative
expansion. The extension of the recent next-to-next-to-leading-order
(NNLO) pQCD calculations for jet production~\cite{prl:118:072002} to
prompt-photon production\footnote{After completion of the work
  presented here, first NNLO calculations for prompt-photon production
  have been completed~\cite{1612.04333}.} will allow a more stringent
test of pQCD. To make such a test with small experimental and
theoretical uncertainties, it is optimal to perform measurements of
prompt-photon production at high photon transverse energies and at the
highest possible centre-of-mass energy of the colliding particles.

Since the dominant production mechanism in $pp$ collisions at the LHC
proceeds via the $qg\rightarrow q\gamma$ process, measurements of
prompt-photon production are sensitive at leading order (LO) to the
gluon density in the
proton~\cite{pr:d30:49,rmp:59:465,np:b297:661,pr:d37:1161,pr:d39:3275,np:b453:334,pr:d73:094007,pr:d82:014015,np:b860:311,epl:101:61002}.
Although prompt photon data were initially included in the
determination of the proton parton distribution functions (PDFs),
their use was abandoned some years ago. Since then, theoretical
developments~\cite{pr:d73:094007,pr:d82:014015} have shown ways to
improve the description of the data in terms of pQCD, and a recent
study quantified the impact of prompt-photon data from hadron
colliders on the gluon density in the proton~\cite{np:b860:311}. New
measurements of prompt-photon production at higher centre-of-mass
energies are expected to further constrain the gluon density in the
proton when combined with previous data.

These measurements can also be used to tune the Monte Carlo (MC)
models to improve the understanding of prompt-photon production. In
addition, precise measurements of these processes aid those searches
for which they are an important background, such as the search for new
phenomena in final states with a photon and missing tranverse
momentum.

Measurements of prompt-photon production at a hadron collider require
isolated photons to avoid the large contribution of photons from
decays of energetic $\pi^0$ and $\eta$ mesons inside jets.  The
production of inclusive isolated photons in $pp$ collisions at
centre-of-mass energies of $\sqrt s = 7$ and $8$~\TeV\ was measured by
the ATLAS~\cite{pr:d83:052005,pl:b706:150,pr:d89:052004,jhep:1608:005} and
CMS~\cite{prl:106:082001,pr:d84:052011} collaborations.

This paper presents measurements of isolated-photon production in $pp$
collisions at $\sqrt s =13$~\TeV\ with the ATLAS detector at the LHC
using a data set with an integrated luminosity of $3.2$~\fb1\
collected during 2015. These measurements are performed in a
phase-space region overlapping with that used in the previous ATLAS
measurement at $\sqrt s = 8$~\TeV~\cite{jhep:1608:005}. Cross sections as
functions of the photon transverse
energy\footnote{\AtlasCoordFootnote} $(\etg)$ are measured in the
range $\etg>125$~\GeV\ for different regions of the photon pseudorapidity
$(\etag)$. The threshold in $\etg$ is chosen so as to avoid the
low-$\etg$ region where both systematic and theoretical uncertainties
increase. Next-to-leading-order (NLO) pQCD and MC event-generator
predictions are compared to the measurements.

\section{The ATLAS detector}
\label{detector}
The ATLAS detector~\cite{PERF-2007-01} is a multi-purpose detector
with a forward-backward symmetric cylindrical geometry. It consists of
an inner tracking detector surrounded by a thin superconducting
solenoid, electromagnetic and hadronic calorimeters, and a muon
spectrometer incorporating three large superconducting toroid
magnets. The inner-detector system is immersed in a $2$~T axial
magnetic field and provides charged-particle tracking in the range
$|\eta|<2.5$. The high-granularity silicon pixel detector is closest
to the interaction region and provides four measurements per track;
the innermost layer, known as the insertable
B-layer~\cite{ATLAS-TDR-19}, was added in 2014 and provides
high-resolution hits at small radius to improve the tracking
performance. The pixel detector is followed by the silicon microstrip
tracker, which typically provides four three-dimensional measurement
points per track. These silicon detectors are complemented by the
transition radiation tracker, which enables radially extended track
reconstruction up to $|\eta|=2.0$. The calorimeter system covers the
range $|\eta|<4.9$. Within the region $|\eta|<3.2$, electromagnetic
calorimetry is provided by barrel and endcap high-granularity
lead/liquid-argon (LAr) electromagnetic calorimeters, with an
additional thin LAr presampler covering $|\eta|<1.8$ to correct for
energy loss in material upstream of the calorimeters; for $|\eta|<2.5$
the LAr calorimeters are divided into three layers in depth. Hadronic
calorimetry is provided by a steel/scintillator-tile calorimeter,
segmented into three barrel structures within $|\eta|<1.7$, and two
copper/LAr hadronic endcap calorimeters, which cover the region $1.5 <
|\eta|<3.2$. The solid angle coverage is completed out to $|\eta|=4.9$
with forward copper/LAr and tungsten/LAr calorimeter modules, which
are optimised for electromagnetic and hadronic measurements,
respectively. Events are selected using a first-level trigger
implemented in custom electronics, which reduces the maximum event
rate of 40~MHz to a design value of 100~kHz using a subset of detector
information. Software algorithms with access to the full detector
information are then used in the high-level trigger to yield a
recorded event rate of about 1~kHz~\cite{trigger:menu}.

\section{Data selection}
\label{datsel}
The data used in this analysis were collected with the ATLAS detector
during the $pp$ collision running period of 2015, when the LHC
operated with a bunch spacing of $25$~ns and a centre-of-mass energy
of $\sqrt s=13$~\TeV. Only events taken in stable beam conditions and
satisfying detector and data-quality requirements are considered. The
total integrated luminosity of the collected sample amounts to 
$3.16 \pm 0.07$~\fb1~\cite{epj:c73:2518,epj:c76:653}. Events were
recorded using a single-photon trigger, with a transverse energy
threshold of $120$~\GeV. The trigger efficiency for isolated photons
with $\etg>125$~\GeV\ and $|\etag|<2.37$, excluding
$1.37<|\etag|<1.56$, is higher than $99\%$.

Events are required to have a reconstructed primary vertex. Primary
vertices are formed from sets of two or more reconstructed tracks,
each with $p_{\mathrm{T}}>400$~\MeV\ and $|\eta|<2.5$, that are mutually
consistent with having originated at the same three-dimensional point
within the luminous region of the colliding proton beams. If multiple
primary vertices are reconstructed, the one with the highest sum of
the $p_{\mathrm{T}}^2$ of the associated tracks is selected as the primary
vertex.

Photon and electron candidates are reconstructed from clusters of
energy deposited in the electromagnetic calorimeter. Candidates
without a matching track or reconstructed conversion
vertex\footnote{Conversion vertex candidates are reconstructed from
  pairs of oppositely charged tracks in the inner detector that are
  likely to be electrons~\cite{epj:c76:666}.} in the inner detector
are classified as unconverted photons~\cite{epj:c76:666}. Those with a
matching reconstructed conversion vertex or a matching track
consistent with originating from a photon conversion are classified as
converted photons. Those matched to a track consistent with
originating from an electron produced in the beam interaction region
are classified as electrons.

The photon identification is based primarily on shower shapes in the
calorimeter~\cite{epj:c76:666}. An initial selection is derived using
the information from the hadronic calorimeter and the lateral shower
shape in the second layer of the electromagnetic calorimeter, where
most of the photon energy is contained. The final tight selection
applies stringent criteria~\cite{epj:c76:666} to these variables,
different for converted and unconverted photon candidates. It also
places requirements on the shower shape in the finely segmented first
calorimeter layer to ensure the compatibility of the measured shower
profile with that originating from a single photon impacting the
calorimeter. When applying the photon identification criteria to
simulated events, corrections are made for small differences in the
average values of the shower-shape variables between data and
simulation. The efficiency of the photon identification varies in the
range $92$--$98$\% for $\etg=125$~\GeV\ and $86$--$98$\% for
$\etg=1$~\TeV, depending on $\etag$ and whether the photon candidate
is classified as unconverted or
converted~\cite{epj:c76:666,ATL-PHYS-PUB-2016-014}. For
$\etg>125$~\GeV, the uncertainty in the photon identification
efficiency varies between $1\%$ and $5\%$, depending on $\etag$ and
$\etg$.

The photon energy measurement is made using calorimeter and tracking
information. A dedicated energy calibration~\cite{epj:c74:3071} is
then applied to the candidates to account for upstream energy loss and
both lateral and longitudinal leakage; a multivariate regression
algorithm to calibrate electron and photon energy measurements was
developed and optimised on simulated events. The  calibration of the
layer energies in the calorimeter is based on the measurement
performed with 2012 data at $\sqrt{s}=8$~\TeV~\cite{epj:c74:3071}. The
overall energy scale in data and the difference in the energy
resolution's constant term\footnote{The relative energy resolution is
  parameterised as $\sigma(E)/E = a/\sqrt{E} \oplus c$, where $a$ is
  the sampling term and $c$ is the constant term.} between data and
simulation are estimated with a sample of $Z$-boson decays to
electrons recorded in 2012 and reprocessed using the same electron
reconstruction and calibration scheme as used for the 2015 data taking
and event processing. The energy scale and resolution corrections are
checked using $Z$-boson decays to electrons recorded in the 2015 data
set. Uncertainties in the measurements performed with this sample are
estimated following a procedure similar to that discussed in
Ref.~\cite{epj:c74:3071}. The difference between the values measured
with the 2015 data and those predicted from the reprocessed 2012 data
is also taken into account in the uncertainties. The uncertainty in
the photon energy scale at high $\etg$ is typically $0.5$--$2.0$\%,
depending on $\etag$. Events with at least one photon candidate with
calibrated $\etg>125$~\GeV\ and $|\etag|<2.37$ are
selected. Candidates in the region $1.37<|\etag|<1.56$, which includes
the transition region between the barrel and endcap calorimeters, are
not considered.

The photon candidate is required to be isolated based on the amount of
transverse energy inside a cone of size $\Delta R=0.4$ in the
$\eta$--$\phi$ plane around the photon candidate, excluding an area of
size $\Delta\eta \times \Delta\phi = 0.125 \times 0.175$ centred on
the photon. The isolation transverse energy is computed from
topological clusters of calorimeter cells~\cite{1603.02934} and is
denoted by $\etiso$. The measured value of $\etiso$ is corrected for
leakage of the photon's energy into the isolation cone and the
estimated contributions from the underlying event (UE) and additional
inelastic $pp$ interactions (pile-up). The latter two corrections are
computed simultaneously on an event-by-event basis~\cite{pl:b706:150}
and the combined correction is typically $2$~\GeV. The combined
correction is computed using a method suggested in
Refs.~\cite{jhep:0804:005,jhep:1004:065}: the $\kt$ jet
algorithm~\cite{pr:d48:3160,np:b406:187} with jet radius $R=0.5$ is
used to reconstruct all jets taking as input topological clusters of
calorimeter cells; no explicit transverse momentum threshold is
applied. The ambient-transverse energy density for the event ($\rho$),
from pile-up and the underlying event, is computed using the median of
the distribution of the ratio between the jet transverse energy and
its area. Finally, $\rho$ is multiplied by the area of the isolation
cone to compute the correction to $\etiso$. In addition, for simulated
events, data-driven corrections to $\etiso$ are applied such that the
peak position in the $\etiso$ distribution coincides in data and
simulation. After all these corrections, $\etiso$ is required to be lower than 
$E_{\mathrm{T,cut}}^{\mathrm{iso}}(\etg)$[\GeV]$\equiv$~\etisocutn~[\GeV]~\cite{jhep:1608:005}. 
The isolation requirement significantly reduces the main background,
which consists of multi-jet events where one jet typically contains a
$\pi^0$ or $\eta$ meson that carries most of the jet energy and is
misidentified as a photon because it decays into an almost collinear
photon pair.

A small fraction of the events contain more than one photon candidate
satisfying the selection criteria. In such events, the highest-$\etg$
(leading) photon is considered for further study. The total number of
data events selected by using the requirements discussed above amounts
to $1\,253\,508$. A summary of the kinematic requirements as well as
the number of selected events in data in each $|\etag|$ region are
included in Table~\ref{tablezero}. The selected sample of events is
used to unfold the distribution in $\etg$ separately for each of the
four regions in $|\etag|$ indicated in Table~\ref{tablezero}; the
unfolding is performed using the samples of MC events described in
Section~\ref{mc} and the results are compared  to the predictions from
the \pyt\ and \sher\ generators as well as to the predictions from NLO
pQCD (see Section~\ref{res}).

\begin{table}[h]
  \caption{
    Kinematic requirements and number of selected events in data for
    each phase-space region.
  }
\setlength{\unitlength}{1.0cm}
\begin{picture} (18.0,3.0)
  \put (0.0,1.0){\scalebox{1.0}{\centerline{
        \begin{tabular}{|l|c|c|c|c|} \hline
   & \multicolumn{4}{c|}{Phase-space region} \\ \hline
Requirement on $\etg$ & \multicolumn{4}{c|}{ $\etg> 125$~\GeV\ } \\ \hline
Isolation requirement & \multicolumn{4}{c|}{  $\etiso<$~\etisocutn~[\GeV] } \\ \hline
Requirement on $|\etag|$ & $|\etag|<0.6$ & $0.6<|\etag|<1.37$ &
$1.56<|\etag|<1.81$ & $1.81<|\etag|<2.37$ \\ \hline
Number of events & $356\,604$ & $480\,466$ & $140\,955$ & $275\,483$ \\ \hline
\end{tabular}
}}}
\end{picture}
\label{tablezero}
\end{table}

\section{Monte Carlo simulations and theoretical predictions}

\subsection{Monte Carlo simulations}
\label{mc}
Samples of MC events were generated to study the characteristics of
signal events. The MC programs \pyt~8.186~\cite{cpc:178:852} and
\sher~2.1.1~\cite{jhep:0902:007} were used to generate the simulated
events. In both generators, the partonic processes were simulated
using tree-level matrix elements, with the inclusion of initial- and
final-state parton showers. Fragmentation into hadrons was performed
using the Lund string model~\cite{prep:97:31} in the case of  \pyt,
and in \sher\ events by a modified version of the cluster
model~\cite{epj:c36:381}. The LO NNPDF2.3~\cite{np:b867:244} PDFs were
used for \pyt\ (NLO CT10~\cite{pr:d82:074024} for \sher) to
parameterise the proton structure. Both samples include a simulation
of the UE. The event-generator parameters were set according to the
``A14'' tune for \pyt~\cite{ATL-PHYS-PUB-2014-021} and the ``CT10''
tune for \sher.  All the samples of generated events were passed
through the {\textsc{Geant4}}-based~\cite{nim:a506:250} ATLAS detector- and
trigger-simulation programs~\cite{epj:c70:823}. They were
reconstructed and analysed by the same program chain as the
data. Pile-up from additional $pp$ collisions  in the same and
neighbouring bunch crossings was simulated by overlaying each MC event
with a variable number of simulated inelastic $pp$ collisions
generated using \pyt 8 with the A2
tune~\cite{ATL-PHYS-PUB-2012-003}. The MC events were weighted to
reproduce the distribution of the average number of interactions per
bunch crossing ($\mu$) observed in the data, referred to as ``pile-up
reweighting''; in this procedure, the $\mu$ value in the data is
divided by a factor of $1.16\pm 0.07$, a rescaling which improves the
agreement between the data and simulation for the observed number of
primary vertices and recovers the fraction of visible cross-section of
inelastic $pp$ collisions as measured in the data~\cite{nc:2:463}.

The \pyt\ simulation of the signal includes LO photon-plus-jet events
from both direct processes (the hard subprocesses 
$qg\rightarrow q\gamma$ and $\qq\rightarrow g\gamma$, called the
``hard'' component) and photon bremsstrahlung in QCD dijet events
(called the ``bremsstrahlung'' component). The \sher\ samples were
generated with LO matrix elements for photon-plus-jet final states
with up to three additional partons ($2\rightarrow n$ processes with
$n$ from $2$ to $5$); the matrix elements were merged with the \sher\
parton shower~\cite{jhep:0905:053} using the ME+PS@LO
prescription. While the bremsstrahlung component was modelled in \pyt\
by final-state QED radiation arising from calculations of all
$2\rightarrow 2$ QCD processes, it was accounted for in \sher\ through
the matrix elements of $2\rightarrow n$ processes with $n\geq 3$; in
the generation of the \sher\ samples, a requirement on the photon
isolation at the matrix-element level was imposed using the criterion
defined in Ref.~\cite{pl:b429:369}.\footnote{This criterion, commonly
  called Frixione's criterion, requires the total transverse energy
  inside a cone of size ${\cal V}$ around the generated final-state
  photon, excluding the photon itself, to be below a certain threshold, 
  $E_{\mathrm{T}}^{\mathrm{max}}({\cal V})=\epsilon\etg ((1-\cos{{\cal V}})/(1-\cos{{\cal R}}))^{n}$,
  for all ${\cal V} < {\cal R}$. The parameters for the threshold were
  chosen to be ${\cal R}=0.3$, $n=2$ and $\epsilon=0.025$.}
 
The predictions of the MC generators at particle level are defined
using those particles with a lifetime $\tau$ longer than $10$~ps;
these particles are referred to as ``stable''. The particles
associated with the overlaid $pp$ collisions (pile-up) are not
considered. The particle-level isolation requirement on the photon was
built summing the transverse energy of all stable particles, except
for muons and neutrinos, in a cone of size $\Delta R=0.4$ around the
photon direction after the contribution from the UE was subtracted;
the same subtraction procedure used on data was applied at the
particle level. Therefore, the cross sections quoted from MC
simulations refer to photons that are isolated by requiring 
$\etiso({\mathrm{particle}})<E_{\mathrm{T,cut}}^{\mathrm{iso}}(\etg)$.

\subsection{Next-to-leading-order pQCD predictions}
\label{nlo}
The NLO pQCD predictions presented in this paper are computed using
the program \jetp~1.3.1\_2~\cite{jhep:0205:028,pr:d73:094007}. 
This program includes a full NLO pQCD calculation of both the direct
and fragmentation contributions to the cross section for the
$pp\rightarrow\gamma\ +\ {\mathrm{X}}$ process.

The number of massless quark flavours is set to five. The
renormalisation scale $\mu_{\mathrm{R}}$ (at which the strong coupling is
evaluated), factorisation scale $\mu_{\mathrm{F}}$ (at which the proton
PDFs are evaluated) and fragmentation scale $\mu_{\mathrm{f}}$ (at which
the fragmentation function is evaluated) are chosen to be
$\mu_{\mathrm{R}}=\mu_{\mathrm{F}}=\mu_{\mathrm{f}}=\etg$. The calculations are
performed using the MMHT2014~\cite{epj:c75:204} parameterisations of
the proton PDFs and the BFG set II of parton-to-photon fragmentation
functions at NLO~\cite{epj:c2:529}. The strong coupling constant is
calculated at two loops with $\alpha_{\mathrm{s}}(m_Z)=0.120$. Predictions
based on other proton PDF sets, namely CT14~\cite{pr:d93:033006} and
NNPDF3.0~\cite{jhep:1504:040}, are also computed. The calculations are
performed using a parton-level isolation criterion which requires the
total transverse energy from the partons inside a cone of size $\Delta
R=0.4$ around the photon direction to be below 
$E_{\mathrm{T,cut}}^{\mathrm{iso}}(\etg)$.

The NLO pQCD predictions refer to the parton level while the
measurements refer to the particle level. Since the data are corrected
for pile-up and UE effects and the distributions are unfolded to a
phase-space definition in which the requirement on $\etiso$ at
particle level is applied after subtraction of the UE, it is expected
that parton-to-hadron corrections to the NLO pQCD predictions are
small. This is confirmed by computing the ratio of the particle-level
cross section for a \pyt\ sample with UE effects to the parton-level
cross section without UE effects:\footnote{The effects of
  hadronisation and UE are also studied separately; the effects of
  including the UE do not cancel those of hadronisation and are
  dominant.} the ratio is consistent with unity within $1\%$ over the
measured range in $\etg$. Therefore, no correction is applied to the
NLO pQCD predictions and an uncertainty of $1\%$ is assigned.

\section{Background estimation and signal extraction}
\label{bacsub}
A non-negligible background contribution remains in the selected
sample, even after imposing the tight identification and isolation
requirements on the photon. This background originates mainly from
multi-jet processes in which a jet is misidentified as a photon.  

The background subtraction relies on a data-driven method based on
signal-suppressed control regions. The background contamination in the
selected sample is estimated using the same two-dimensional sideband
technique as in the previous
analyses~\cite{pr:d83:052005,pl:b706:150,pr:d85:092014,jhep:1608:005,np:b875:483}
and then subtracted bin-by-bin from the observed yield. In this
method, the photon is classified as:
\begin{itemize}
\item ``isolated'', if $\etiso<E_{\mathrm{T,cut}}^{\mathrm{iso}}(\etg)$;
\item ``non-isolated'', if $\etiso> E_{\mathrm{T,cut}}^{\mathrm{iso}}(\etg) +2\ {\mathrm{\GeV}}$ 
  and $\etiso< 50$~\GeV;
\item ``tight'', if it satisfies the tight photon identification
  criteria;
\item ``non-tight'', if it fails at least one of four tight
  requirements on the shower-shape variables computed from the energy
  deposits in the first layer of the electromagnetic calorimeter, but
  satisfies the tight requirement on the total lateral shower width in
  the first layer and all the other tight identification
  criteria~\cite{epj:c76:666}.
\end{itemize}

In the two-dimensional plane formed by $\etiso$ and the photon
identification variables, which are chosen because they are expected
to be independent for the background, four regions are defined:
\begin{itemize}
\item{$A$}: the ``signal'' region, containing tight isolated photon 
  candidates;
\item{$B$}: the ``non-isolated'' background control region, containing
  tight non-isolated photon candidates;
\item{$C$}: the ``non-tight'' background control region,
  containing isolated non-tight photon candidates;
\item{$D$}: the background control region containing non-isolated 
  non-tight photon candidates.
\end{itemize}

The signal yield $N_A^{\mathrm{sig}}$ in region $A$ is estimated by using
the relation

\begin{equation}
  N_A^{\mathrm{sig}}=N_A-R^{\mathrm{bg}}\cdot(N_B-f_B N_A^{\mathrm{sig}})\cdot
  \frac{(N_C-f_C N_A^{\mathrm{sig}})}{(N_D-f_D N_A^{\mathrm{sig}})},
  \label{eqone}
  \end{equation}
where $N_K$, with $K=A,B,C,D$, is the number of events in region $K$
and 
$R^{\mathrm{bg}}=N_A^{\mathrm{bg}}\cdot N_D^{\mathrm{bg}}/(N_B^{\mathrm{bg}}\cdot N_C^{\mathrm{bg}})$
is the so-called background correlation and is taken as $R^{\mathrm{bg}}=1$
for the nominal results; $N_K^{\mathrm{bg}}$ with $K=A,B,C,D$ is the number
of background events in each region. Equation~(\ref{eqone}) takes into
account the expected number of signal events in the three background
control regions ($N_K^{\mathrm{sig}}$) via the signal leakage fractions,
$f_K=N_K^{\mathrm{sig}}/N_A^{\mathrm{sig}}$ with $K=B,C,D$, which are estimated
using the MC simulations of the signal. A systematic uncertainty is
assigned to the modelling of the signal leakage fractions (see
Section~\ref{syst}). The only assumption underlying Eq.~(\ref{eqone})
is that the isolation and identification variables are independent for
background events, thus $R^{\mathrm{bg}}=1$. This assumption is verified
both in simulated background samples and in data in a
background-dominated region~\cite{jhep:1608:005}.  A study of
$R^{\mathrm{bg}}$ in background-dominated regions, accounting for
signal leakage using either the \pyt\ or \sher\ simulations, shows deviations from
unity which are then propagated through Equation~(\ref{eqone}) and
taken as systematic uncertainties. The signal purity, defined as
$N_A^{\mathrm{sig}}/N_A$, is above $90\%$ for $\etg=125$~\GeV\ in all
$\etag$ regions and increases as $\etg$ increases. The signal purity
is similar whether \pyt\ or \sher\ is used to extract the signal
leakage fractions and the difference is taken as a systematic
uncertainty.

There is an additional background from electrons misidentified as
photons, mainly produced in Drell-Yan $Z^{(*)}/\gamma^*\rightarrow e^+e^-$ and 
$W^{(*)}\rightarrow e\nu$ processes. Such misidentified electrons are
largely suppressed by the photon selection. The remaining electron
background is estimated using MC techniques and found to be negligible
in the phase-space region of the analysis presented here.

\section{Unfolding}
\label{cor}
The isolated-photon cross section is measured as a function of $\etg$
in different regions of $|\etag|$. The phase-space regions are listed
in Table~\ref{tablezero}. The data distributions, after background
subtraction, are unfolded to the particle level using bin-by-bin
correction factors determined using the MC samples. These correction
factors take into account the efficiency of the selection criteria and
the purity and efficiency of the photon reconstruction. The data
distributions are unfolded to the particle level via the formula

\begin{equation}
  \frac{{\mathrm{d}}\sigma}{{\mathrm{d}}\etg}(i)=\frac{N_A^{\mathrm{sig}}(i)\ C^{\mathrm{MC}}(i)}{{\cal L}\
    \Delta \etg(i)},
  \label{eqthree}
\end{equation}
where $({\mathrm{d}}\sigma/{\mathrm{d}}\etg)(i)$ is the cross section as a
function of the observable $\etg$ in bin $i$, $N_A^{\mathrm{sig}}(i)$ is
the number of background-subtracted data events in bin $i$, 
$C^{\mathrm{MC}}(i)$ is the correction factor in bin $i$, ${\cal L}$ is the
integrated luminosity and $\Delta \etg(i)$ is the width of bin
$i$. The correction factors are computed using the MC samples of
events as $C^{\mathrm{MC}}(i)=N^{\mathrm{MC}}_{\mathrm{part}}(i)/N^{\mathrm{MC}}_{\mathrm{reco}}(i)$,
where $N^{\mathrm{MC}}_{\mathrm{part}}(i)$ is the number of events which satisfy
the kinematic constraints of the phase-space region at the particle
level, and $N^{\mathrm{MC}}_{\mathrm{reco}}(i)$ is the number of events which
meet all the selection criteria at the reconstruction level.

The nominal cross sections are measured using the correction factors
from \pyt\ and the deviations from these results when using \sher\ to
unfold the data are taken to represent systematic uncertainties in how
the parton-shower and hadronisation models affect the corrections. 
The correction factors increase as $\etg$ increases and vary between
$1.04$ and $1.24$ depending on $\etg$ and $\etag$. The results of the
bin-by-bin unfolding procedure are checked with a Bayesian unfolding
method~\cite{nim:a362:487}, giving consistent results.

\section{Experimental and theoretical uncertainties}

\subsection{Experimental uncertainties}
\label{syst}
The primary sources of systematic uncertainty that affect the
measurements are investigated. These sources include photon
identification,  photon energy scale and resolution, background
subtraction, modelling of the final state, pile-up, MC sample
statistics, trigger and luminosity.

\begin{itemize}
\item {\textbf{Photon identification efficiency.}}
  The uncertainty in the photon identification efficiency is estimated
  from the effect of differences between shower-shape variable
  distributions in data and simulation. From the studies presented in
  Ref.~\cite{epj:c76:666}, this procedure is found to provide a
  conservative estimate of the uncertainties.\footnote{The photon
    identification efficiencies from data-driven methods and MC
    simulations were compared in Ref.~\cite{epj:c76:666}. No significant
    difference is observed between the data-driven measurements and the
    nominal or corrected (for the small differences in the average
    values of the shower-shape variables between data and simulation)
    simulation for $\etg>60$~\GeV.} The resulting uncertainty in the
measured cross sections increases from $1$--$2$\% at $\etg=125$~\GeV\
to $2$--$6$\% at $\etg\sim 1$~\TeV.

\item {\textbf{Photon energy scale and resolution.}}
  A detailed assessment of the uncertainties in the photon energy scale
  and resolution is made using the same method developed with $8$~\TeV\
  data~\cite{epj:c74:3071}. The sources of uncertainty include: the
  uncertainty in the overall energy scale adjustment using 
  $Z\rightarrow e^+e^-$; the uncertainty in the non-linearity of the
  energy measurement at the cell level; the uncertainty in the relative
  calibration of the different calorimeter layers; the uncertainty in
  the amount of material in front of the calorimeter; the uncertainty in
  the modelling of the reconstruction of photon conversions; the
  uncertainty in the modelling of the lateral shower shape; the
  uncertainty in the modelling of the sampling term; the uncertainty in
  the measurement of the constant term in $Z$-boson decays. Additional
  systematic uncertainties are included to take into account the
  differences between the 2012 and 2015 configurations. These
  uncertainties are modelled using independent components to account for
  their $\eta$ dependence. All the components are propagated through the
  analysis separately to maintain the full information about the
  correlations.\\
  The systematic uncertainties in the measured cross sections due to the
  effects mentioned above are estimated by varying by $\pm 1\sigma$ each
  individual source of uncertainty separately in the MC simulations and
  then added in quadrature. The resulting uncertainty increases from
  about $2\%$ at $\etg=125$~\GeV\ to about $5\%$ at $\etg\sim 1$~\TeV\
  except in the $1.56<|\etag|<1.81$ region, where it increases from
  about $7\%$ at $\etg=125$~\GeV\ to about $18\%$ at $\etg\sim 1$~\TeV.
 
\item {\textbf{Definition of the background control regions.}}
  The estimation of the background contamination in the signal region is
  affected by the choice of background control regions. The control
  regions $B$ and $D$ are defined by the lower and upper limits on
  $\etiso$ and the choice of inverted photon identification variables
  used in the selection of non-tight photons. To study the dependence on
  the specific choices, these definitions are varied over a wide
  range. The lower limit on $\etiso$ in regions $B$ and $D$ is varied by
  $\pm 1$~\GeV, which is larger than any difference between data and
  simulations and still provides a sufficient sample to perform the
  data-driven subtraction. The upper limit on $\etiso$ in regions $B$
  and $D$ is removed. The resulting uncertainty in the measured cross
  sections is negligible.\\ 
  Likewise, the choice of inverted photon identification variables is
  varied. The analysis is repeated using different sets of variables:
  tighter (looser) identification criteria are defined by applying tight
  requirements to an extended (restricted) set of shower-shape variables
  in the first calorimeter layer. The resulting uncertainty in the
  measured cross sections is typically smaller than $2\%$.

\item {\textbf{Photon identification and isolation correlation in the
    background.}}
  The photon isolation and identification variables used to define the
  plane in the two-dimensional sideband method to subtract the
  background are assumed to be independent for background events
  ($R^{\mathrm{bg}}=1$ in Eq.~(\ref{eqone})). Any correlation between these
  variables affects the estimation of the purity of the signal and leads
  to systematic uncertainties in the background-subtraction procedure. A
  range in $R^{\mathrm{bg}}$ is set to cover the deviations from unity
  observed for the estimations based on subtracting the signal leakage
  with either \pyt\ or \sher\ MC samples. The resulting range in
  $R^{\mathrm{bg}}$, which is taken as the uncertainty, is
  $0.8<R^{\mathrm{bg}}<1.2$ for $0.6<|\etag|<1.37$ and $1.81<|\etag|<2.37$; 
  for the region $|\etag|<0.6$ ($1.56<|\etag|<1.81$), the range is 
  $0.8<R^{\mathrm{bg}}<1.2$ ($0.75<R^{\mathrm{bg}}<1.25$) at low $\etg$ and
  increases to $0.65<R^{\mathrm{bg}}<1.35$ ($0.6<R^{\mathrm{bg}}<1.4$) at high
  $\etg$. The resulting uncertainty in the measured cross sections is
  typically smaller than $2\%$.

\item {\textbf{Parton-shower and hadronisation model dependence.}}
  The effects due to the parton-shower and hadronisation models in the
  signal purity and correction factors are studied separately; the
  effects are estimated as the differences observed between the nominal
  results and those obtained using \sher\ MC samples either for the
  determination of the signal leakage fractions or the unfolding
  correction factors. The resulting uncertainties in the measured cross
  sections are typically smaller than $2\%$.

\item {\textbf{Photon isolation modelling.}}
  The differences between the nominal results and those obtained without
  applying the data-driven corrections to $\etiso$ in simulated events
  are taken as systematic uncertainties in the measurements due to the
  modelling of $\etiso$ in the MC simulation. The resulting uncertainty
  in the measured cross sections is smaller than $2\%$.

\item {\textbf{Signal modelling.}}
  The MC simulation of the signal is used to estimate the signal leakage
  fractions in the two-dimensional sideband method for background
  subtraction and to compute the bin-by-bin correction factors. The
  \pyt\ simulation is used with the mixture of the hard and
  bremsstrahlung components as predicted by the generator to yield the
  background-subtracted data distributions and to compute the correction
  factors; in the predicted mixture, the relative contribution of the
  bremsstrahlung component amounts to $\approx 30\%$. The uncertainty
  related to the simulation of the hard and bremsstrahlung components is
  estimated by performing the background subtraction and the calculation
  of the correction factors using a mixture with either two or zero
  times the amount of the bremsstrahlung component. The resulting
  uncertainty in the measured cross sections is typically smaller than $1\%$.
 
\item {\textbf{Pile-up.}}
  The uncertainty is estimated by changing the nominal rescaling factor
  of $1.16$ from $1.09$ to $1.23$ and re-evaluating the reweighting
  factors. The resulting uncertainty in the measured cross sections is
  typically smaller than $0.5\%$.

\end{itemize}

The total systematic uncertainty is computed by adding in quadrature
the uncertainties from the sources listed above and the statistical
uncertainty of the MC samples as well as the uncertainty in the
trigger efficiency. The uncertainty in the integrated luminosity is
$2.1\%$~\cite{epj:c76:653}. This uncertainty is fully correlated in
all bins of all the measured cross sections and is shown
separately. The total systematic uncertainty is smaller than $5\%$ for
$|\etag|<1.37$. For $1.56<|\etag|<1.81$ ($1.81<|\etag|<2.37$), it
increases from $\approx 8\%\ (4\%)$ at $\etg=125$~\GeV\ to 
$\approx 19\%\ (11\%)$ at the high end of the spectrum. For
$\etg\lesssim 600$~\GeV, the systematic uncertainty dominates the
total experimental uncertainty, while for higher $\etg$ values, the
statistical uncertainty of the data limits the precision of the
measurements.

\subsection{Theoretical uncertainties}
The following sources of uncertainty in the theoretical predictions
are considered:
\begin{itemize}
\item The uncertainty in the NLO pQCD predictions due to terms beyond
  NLO is estimated by repeating the calculations using values of
  $\mu_{\mathrm{R}}$, $\mu_{\mathrm{F}}$ and $\mu_{\mathrm{f}}$ scaled by the factors
  $0.5$ and $2$. The three scales are either varied simultaneously,
  individually or by fixing one and varying the other two. In all
  cases, the condition $0.5\leq\mu_A/\mu_B\leq 2$ is imposed, where
  $A,B={\mathrm{R,\ F,\ f}}$ and $A\neq B$. The final uncertainty is taken
  as the largest deviation from the nominal value among the 14
  possible variations.
\item The uncertainty in the NLO pQCD predictions due to imperfect
  knowledge of the proton PDFs is estimated by repeating the
  calculations using the $50$ sets from the MMHT2014 error
  analysis~\cite{epj:c75:204} and applying the Hessian
  method~\cite{pr:d65:014013,0110378} for evaluation of the PDF
  uncertainties.
\item The uncertainty in the NLO pQCD predictions due to that in the
  value of $\alpha_{\mathrm{s}}(m_Z)$ is estimated by repeating the
  calculations using two additional sets of proton PDFs from the
  MMHT2014 analysis, for which different values of 
  $\alpha_{\mathrm{s}}(m_Z)$ were assumed in the fits, namely 
  $\alpha_{\mathrm{s}}(m_Z)=0.118$ and $0.122$; in this way, the
  correlation between $\alpha_{\mathrm{s}}$ and the PDFs is preserved.
\item An uncertainty of $1\%$ is assigned due to the non-perturbative
  effects of hadronisation and UE (see Section~\ref{nlo}).
\end{itemize}

The dominant theoretical uncertainty is that arising from the terms
beyond NLO and amounts to $10$--$15$\% for all $\etag$ regions. The
uncertainty arising from those in the PDFs increases from $1\%$ at
$\etg=125$~\GeV\ to $3$--$4$\% at high $\etg$. The uncertainty arising
from the value of $\alpha_{\mathrm{s}}(m_Z)$ is below $2\%$. The total
theoretical uncertainty is obtained by adding in quadrature the
individual uncertainties listed above and amounts to $10$--$15$\%.

\section{Results}
\label{res}
\begin{figure}[p]
  \setlength{\unitlength}{1.0cm}
  \begin{picture} (18.0,15.0)
    \put (0.0,8.0){\includegraphics[width=9cm,height=9cm]{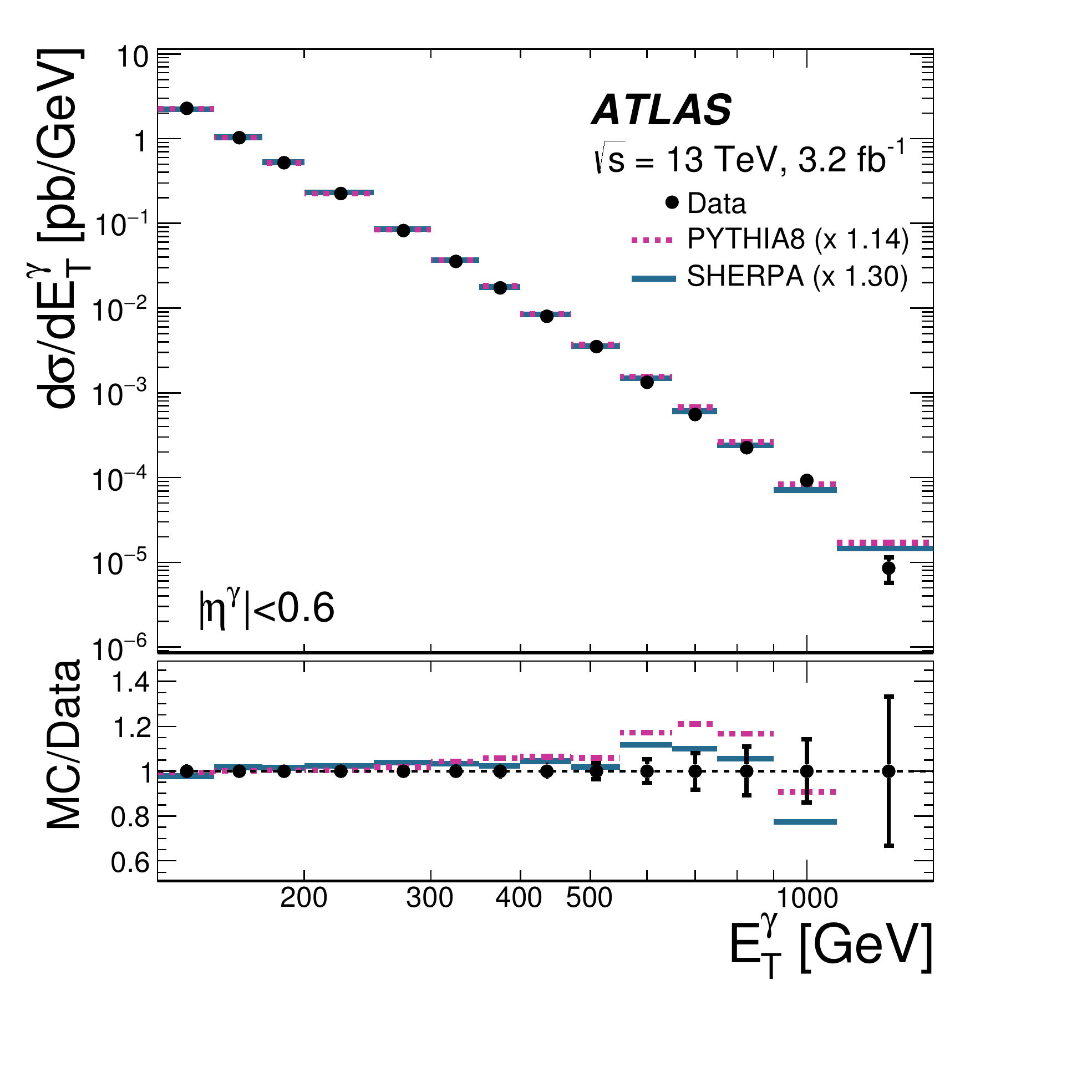}}
    \put (8.0,8.0){\includegraphics[width=9cm,height=9cm]{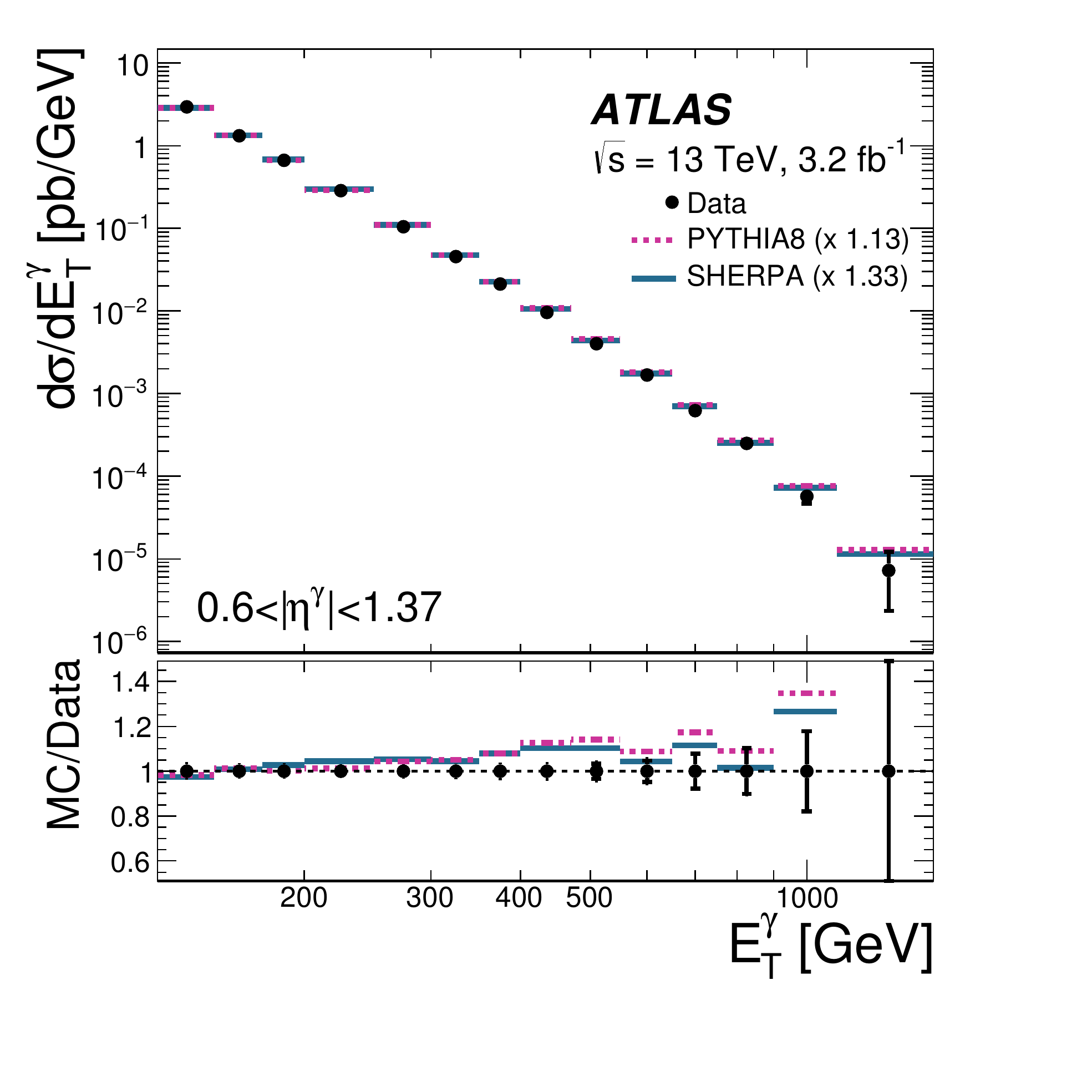}}
    \put (0.0,0.0){\includegraphics[width=9cm,height=9cm]{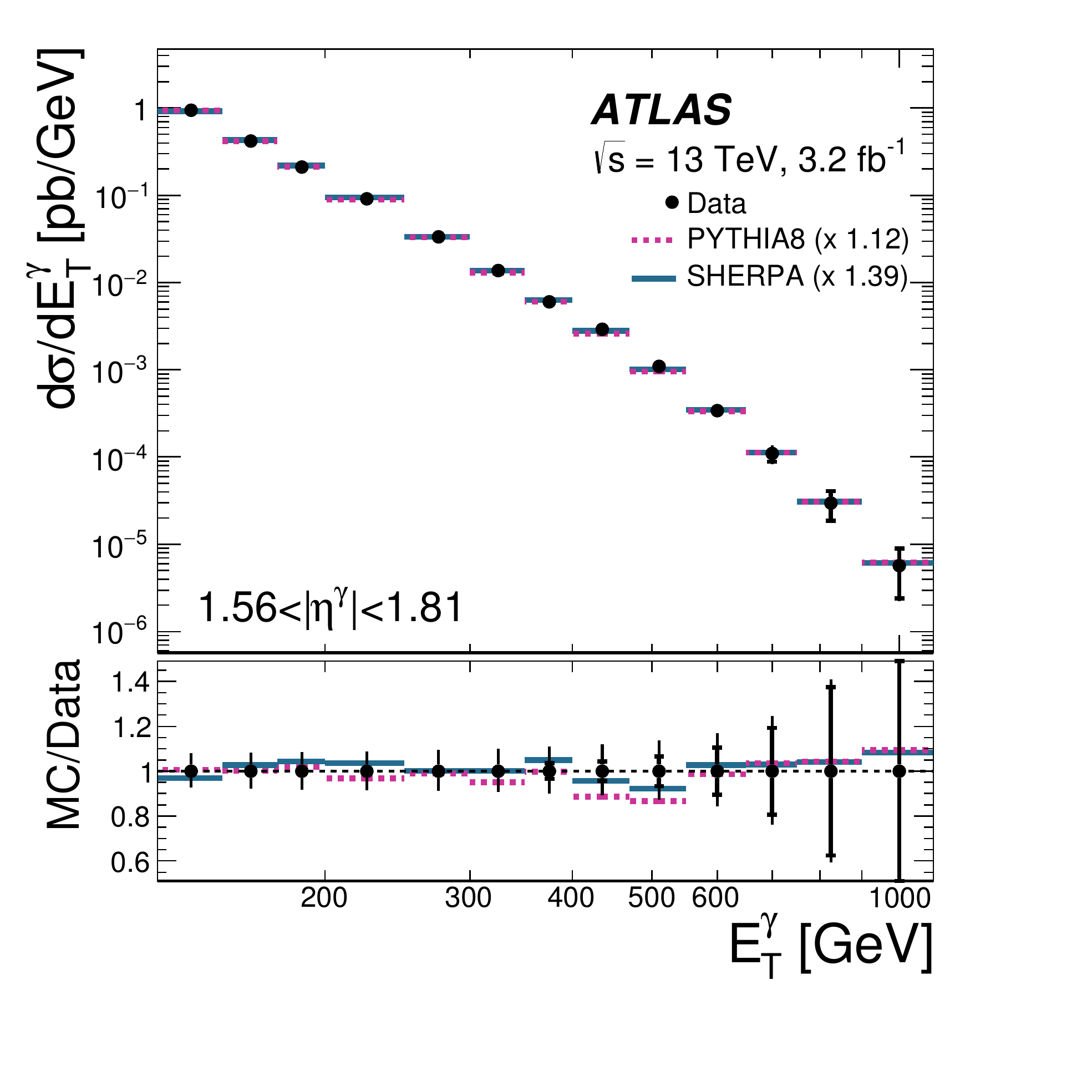}}
    \put (8.0,0.0){\includegraphics[width=9cm,height=9cm]{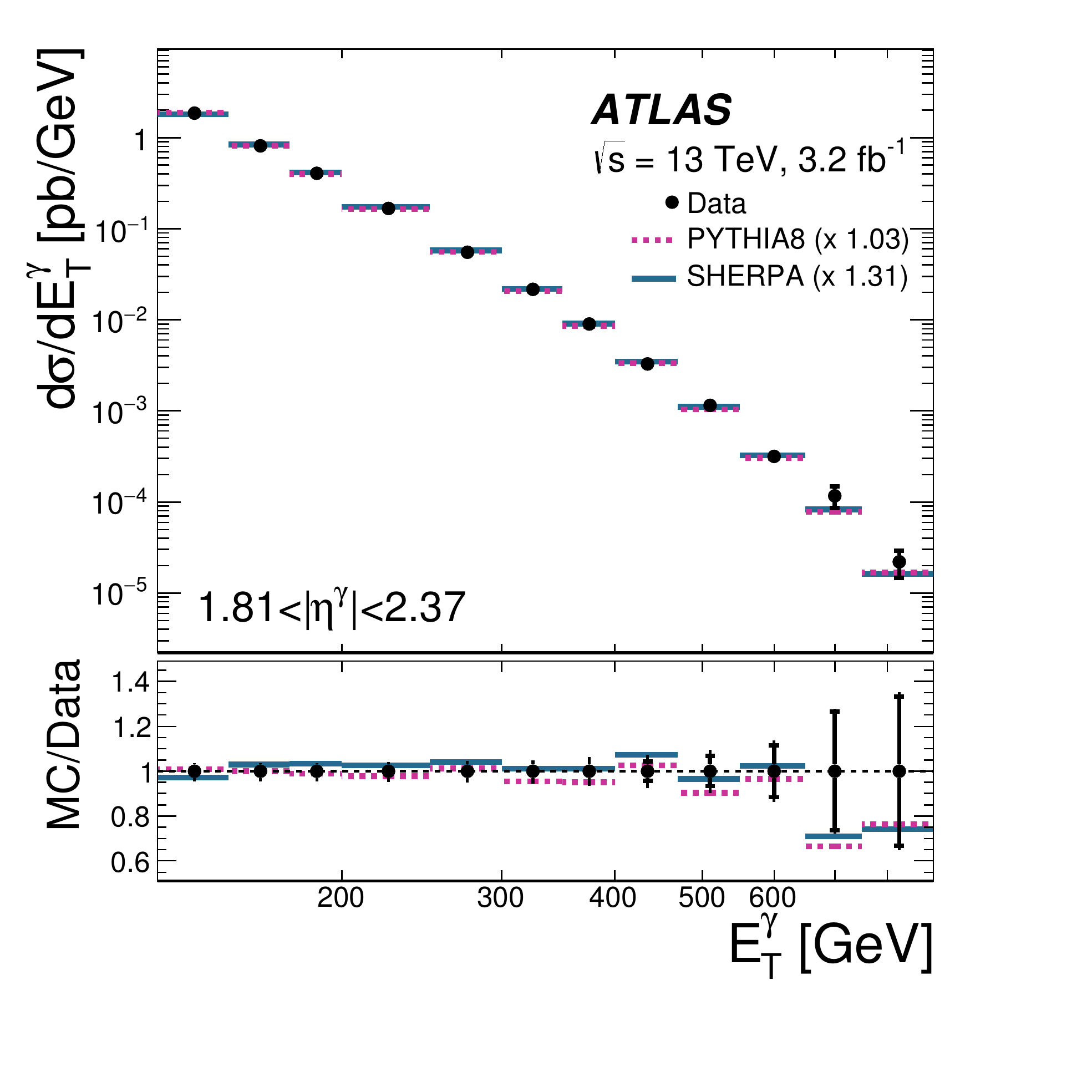}}
    \put (3.7,9.0){{\textbf{\small (a)}}}
    \put (11.7,9.0){{\textbf{\small (b)}}}
    \put (3.7,1.0){{\textbf{\small (c)}}}
    \put (11.7,1.0){{\textbf{\small (d)}}}
  \end{picture}
  \caption
  {
  Measured cross sections for isolated-photon production (dots) as
  functions of $\etg$ in (a) $|\etag|<0.6$, (b) $0.6<|\etag|<1.37$,
  (c) $1.56<|\etag|<1.81$ and (d) $1.81<|\etag|<2.37$. The predictions
  from \pyt\ (dashed lines) and \sher\ (solid lines) are also shown;
  these predictions are normalised to the measured integrated cross
  section in each region of $|\etag|$ using the values indicated in
  parentheses. The bottom part of each figure shows the ratio of the
  MC predictions to the measured cross section. The inner (outer)
  error bars represent the statistical uncertainties (the statistical
  and systematic uncertainties, excluding that on the luminosity,
  added in quadrature). For most of the points, the inner error bars
  are smaller than the marker size and, thus, not visible.
}
\label{figres1}
\end{figure}

Figure~\ref{figres1} shows the isolated-photon cross section as a
function of $\etg$ in four different regions of $\etag$. The measured
cross sections decrease by approximately five orders of magnitude in
the measured range. Values of $\etg$ up to $1.5$~\TeV\ are
accessed. The cross-section distributions measured in the four
different regions of $\etag$ have similar shapes.

The predictions of the \pyt\ and \sher\ MC models are compared to the
measurements in Figure ~\ref{figres1}. These predictions are
normalised to the measured integrated cross section in each $\etag$
region. The difference in normalisation between data and \pyt\ (\sher)
is $\sim +10\%\ (+30\%)$ and attributed to the fact that these
generators are based on tree-level matrix elements, which are affected
by a large normalisation uncertainty due to missing higher-order
terms. The predictions of both \pyt\ and \sher\ give a good
description of the shape of the measured cross-section distributions
for $\etg\lesssim 500$~\GeV\ in the range $|\etag|<1.37$ and in the
whole measured $\etg$ range for $1.56<|\etag|<2.37$.

\begin{figure}[h]
 \setlength{\unitlength}{1.0cm}
\begin{picture} (18.0,15.0)
\put (0.0,0.0){\centerline{\includegraphics[width=15cm,height=15cm]{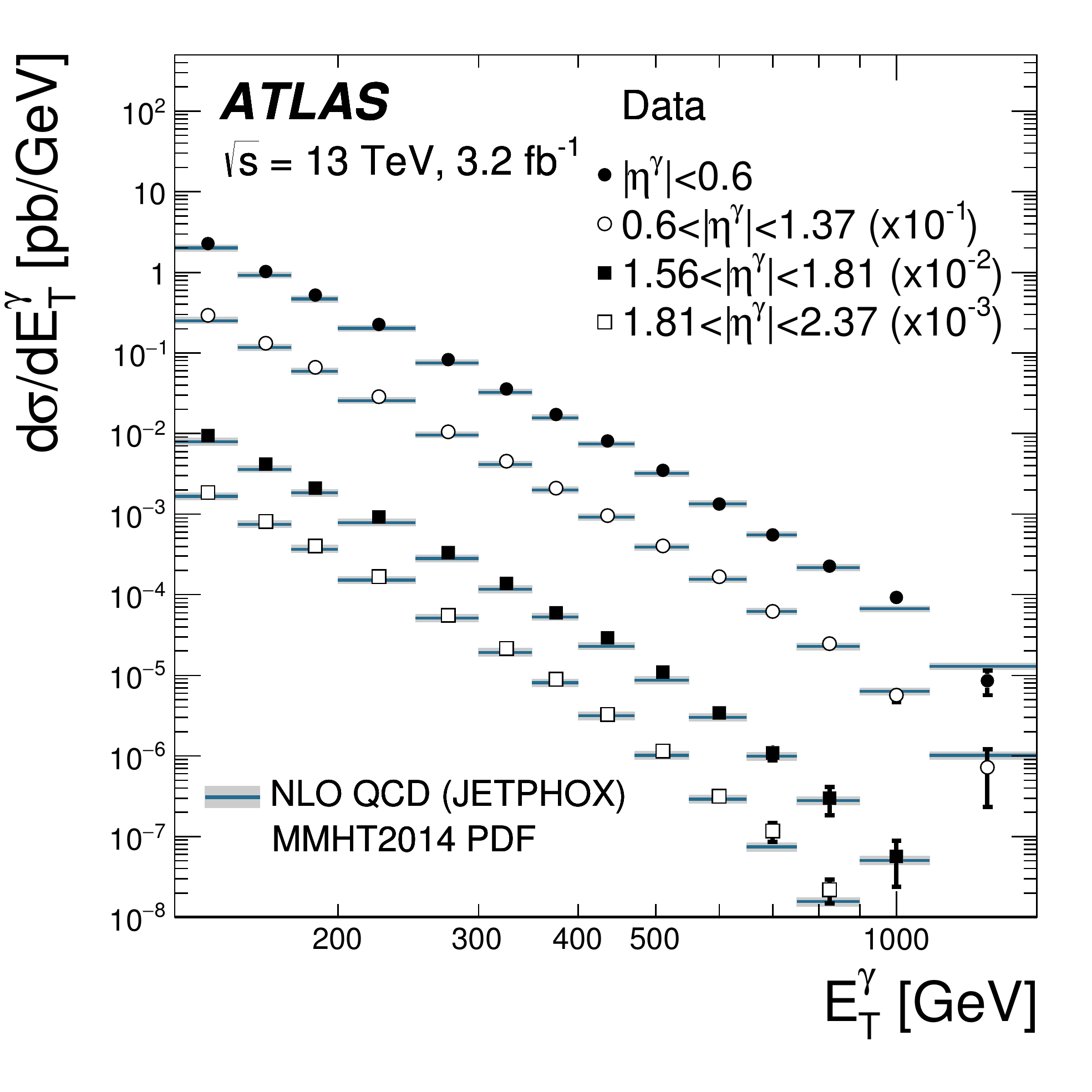}}}
\end{picture}
\caption
{
  Measured cross sections for isolated-photon production as functions
  of $\etg$ in $|\etag|<0.6$ (black dots), $0.6<|\etag|<1.37$ (open
  circles), $1.56<|\etag|<1.81$ (black squares) and
  $1.81<|\etag|<2.37$ (open squares). The NLO pQCD predictions from
  \jetp\ based on the MMHT2014 PDFs (solid lines) are also
  shown. The measurements and the predictions are normalised by the
  factors shown in parentheses to aid visibility. The error bars
  represent the statistical and systematic uncertainties added in
  quadrature. The shaded bands display the theoretical uncertainty.
}
\label{figres2}
\end{figure}

\begin{figure}[p]
\setlength{\unitlength}{1.0cm}
\begin{picture} (18.0,17.0)
\put (0.0,0.0){\centerline{\includegraphics[width=17cm,height=17cm]{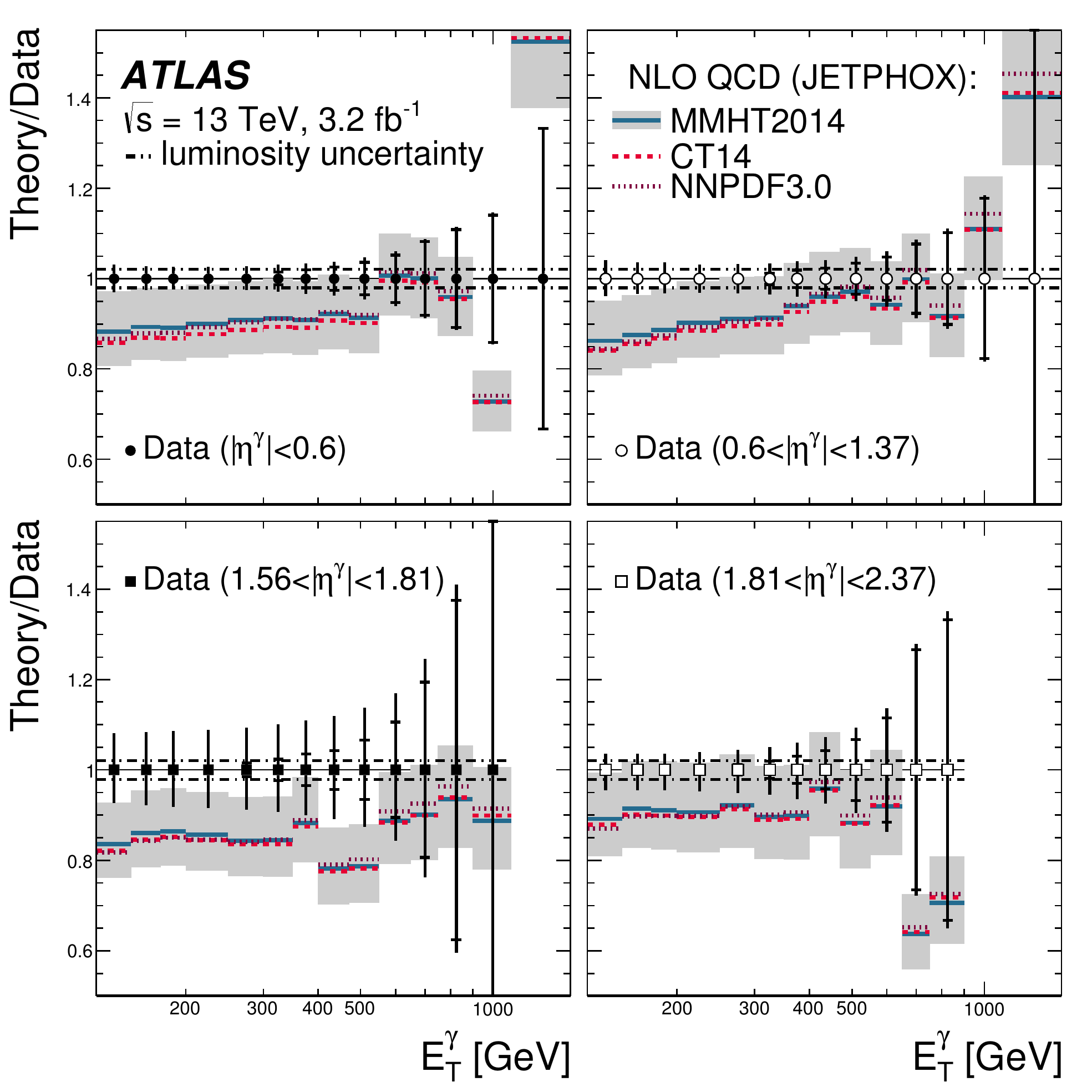}}}
\put (7.5,9.4){{\textbf{\small (a)}}}
\put (15.1,9.4){{\textbf{\small (b)}}}
\put (7.5,1.8){{\textbf{\small (c)}}}
\put (15.1,1.8){{\textbf{\small (d)}}}
\end{picture}
\caption
{
  Ratio of the NLO pQCD predictions from \jetp\ based on the
  MMHT2014 PDFs to the measured cross sections for isolated-photon
  production (solid lines)  as a function of $\etg$ in (a)
  $|\etag|<0.6$, (b) $0.6<|\etag|<1.37$, (c) $1.56<|\etag|<1.81$ and
  (d) $1.81<|\etag|<2.37$. The inner (outer) error bars represent the
  statistical uncertainties (statistical and systematic uncertainties,
  excluding that on the luminosity, added in quadrature), the
  dot-dot-dashed lines represent the uncertainty due to the luminosity
  measurement and the shaded bands display the theoretical uncertainty
  of the calculation based on the MMHT2014 PDFs. The ratio of the NLO
  pQCD predictions based on the CT14 (dashed lines) or NNPDF3.0
  (dotted lines) PDF sets to the data are also included.
}
\label{figres3}
\end{figure}

Figure~\ref{figres2} shows the measured isolated-photon cross sections
as functions of $\etg$ in four different regions of $\etag$ compared
with the predictions of the NLO pQCD calculations of \jetp\
based on the MMHT2014 proton PDF set. The ratios of the theoretical
predictions based on different PDF sets to the measured cross sections
are shown in Figure~\ref{figres3}. The predictions based on MMHT2014,
CT14 and NNPDF3.0 are very similar, the differences being much smaller
than the theoretical scale uncertainties. For most of the points, the
theoretical uncertainties are larger than those of experimental
origin. Differences are observed between data and the predictions of
up to $10$--$15$\% depending on $\etg$ and $|\etag|$; since the
theoretical uncertainties are $10$--$15$\% and cover those
differences, it is concluded that the NLO pQCD predictions provide an
adequate description of the measurements.

The measured cross sections are larger than those at 
$\sqrt s = 8$~\TeV~\cite{jhep:1608:005} by approximately a factor of two at
low $\etg$ ($\etg\sim 125$~\GeV) and by approximately an order of
magnitude at the high end of the spectrum in each region of
$|\etag|$. Such increases in the measured cross section are expected
from the increase in the centre-of-mass energy. The experimental
uncertainties of the measurements at $\sqrt s = 8$ and $13$~\TeV\ are
comparable. For both centre-of-mass energies the NLO theoretical
uncertainties are of similar size and comparable to the differences
between the predictions and the data; since, in addition, the
experimental uncertainties are smaller than those differences, the
inclusion of NNLO pQCD corrections might improve the description of
the two sets of measurements.

The measured fiducial cross section for inclusive isolated-photon
production in the phase-space region given by $\etg>125$~\GeV\ and
$|\etag|<2.37$ (excluding the region of $1.37<|\etag|<1.56$) and
isolation $\etiso<E_{\mathrm{T,cut}}^{\mathrm{iso}}(\etg)$ is 
$$\sigma_{\mathrm{meas}}=399 \pm 13\ ({\mathrm{exp.}})\ \pm 8\ ({\mathrm{lumi.}})\ {\mathrm{pb}},$$ 
where ``exp.'' denotes the sum in quadrature of the statistical and
systematic uncertainties and ``lumi.'' denotes the uncertainty due to
that in the integrated luminosity, details of which are listed in
Table~\ref{tableint}.

The fiducial cross section predicted at NLO in pQCD by \jetp\
using the MMHT2014 PDFs is 
$$\sigma_{\mathrm{NLO}}=352
_{-29}^{+36}\ ({\mathrm{scale}})\ \pm 3 \ ({\mathrm{PDF}})\
\pm 6 \ ({\mathrm{\alpha_{\mathrm{s}}}})\ \pm 4 \ ({\mathrm{non\mhyphen perturb.}})\ {\mathrm{pb}},$$
which is $12\%$ lower than the measurement, but consistent within the
experimental and theoretical uncertainties.

\FloatBarrier
\begin{table}[h]
\caption{
 Uncertainties (in pb) in the fiducial cross section: photon
 identification (``$\gamma$ ID''), photon energy scale and resolution
 (``$\gamma$ ES+ER''), lower limit in $\etiso$ in regions $B$ and $D$
 (``$\etiso$ Gap''), removal of upper limit in $\etiso$ in regions $B$
 and $D$ (``$\etiso$ upp. lim.''), variation of the inverted photon
 identification variables (``$\gamma$ invert. var.''), correlation
 between $\gamma$ ID and isolation in the background (``$R^{\mathrm{
   bg}}$''), signal leakage fractions of \sher\ (``Leak. \sher''), unfolding with \sher\ (``Unf. \sher''),
 modelling of $\etiso$ in MC simulation (``$\etiso$ MC''), mixture of
 hard and bremsstrahlung components in MC samples (``Hard and brem''),
 pile-up (``Pile-up''), statistical uncertainty in MC samples (``MC
 stat.''), trigger (``Trigger''), statistical uncertainty in data
 (``Data stat.'') and luminosity (``Luminosity'').
}
\setlength{\unitlength}{1.0cm}
\begin{picture} (18.0,3.0)
\put (0.0,1.0){\centerline{\scalebox{1.0}{
\begin{tabular}{|l|c|l|c|l|c|} \hline
 \multicolumn{6}{|c|}{Uncertainties [pb]} \\ \hline
$\gamma$ ID            & $(-5.2,+5.4)$     & $\gamma$ ES+ER        & $(-7.9,+8.4)$ 
                                                                       & $\etiso$ Gap                & $\pm 0.3$ \\
$\etiso$ upp. lim.     & $+0.6$                        & $\gamma$ invert. var. & $(-4.1,+3.5)$
                                                                       & $R^{\mathrm{bg}}$             & $(-6.2,+6.1)$ \\
Leak. \sher\ & $\pm 4.1$                   & Unf. \sher\      & $\pm 2.9$ 
                                                                       & $\etiso$ MC                 & $-2.8$ \\
Hard and brem         & $(-1.0,+1.9)$      & Pile-up                        & $(-1.1,+1.3)$ 
                                                                        & MC stat.                     & $\pm 0.4$ \\
Trigger                     & $\pm 1.1$                    & Data stat.                   & $\pm 0.4$
                                                                        & Luminosity                 & $\pm 8.4$  \\ \hline
\end{tabular}
}}}
\end{picture}
\label{tableint}
\end{table}

\section{Summary}
\label{conc}
A measurement of the cross section for inclusive isolated-photon
production in $pp$ collisions at $\sqrt s=13$~\TeV\ with the ATLAS
detector at the LHC is presented using a data set with an integrated
luminosity of $3.2$~\fb1. Cross sections as functions of $\etg$ are
measured in four different regions of $\etag$ for photons with
$\etg>125$~\GeV\ and $|\etag|<2.37$, excluding the region
$1.37<|\etag|<1.56$. Selection of isolated photons is ensured by
requiring that the transverse energy in a cone of size $\Delta R=0.4$
around the photon is smaller than \etisocutn~[\GeV]. Values of $\etg$
up to $1.5$~\TeV\ are measured. The fiducial cross section is measured
to be 
$\sigma_{\mathrm{meas}}=399 \pm 13\ ({\mathrm{exp.}})\ \pm 8\ ({\mathrm{lumi.}})\ {\mathrm{pb}}$.

The experimental systematic uncertainties are evaluated such that the
correlations with previous ATLAS measurements of prompt-photon
production can be used in the fits of the proton parton distribution
functions. A combined fit at NNLO pQCD of the measurements in $pp$
collisions at centre-of-mass energies of $8$ and $13$~\TeV\ which
takes into account the correlated systematic uncertainties has the
potential to constrain further the proton PDFs than either set of
measurements alone.

The predictions of the  \pyt\ and \sher\ Monte Carlo models give a
good description of the shape of the measured cross-section
distributions except for $\etg\gtrsim 500$~\GeV\ in the regions
$|\etag|<0.6$ and $0.6<|\etag|<1.37$. The next-to-leading-order pQCD
predictions, using \jetp\ and based on different sets of proton
PDFs, provide an adequate description of the data within the
experimental and theoretical uncertainties. For most of the phase
space the theoretical uncertainties are larger than those of
experimental nature and dominated by the terms beyond NLO, from which
it is concluded that NNLO pQCD corrections are needed to make an even
more stringent test of the theory.

\section*{Acknowledgements}

% Acknowledgements for papers with collision data
% Version 6-Mar-2017

% Standard acknowledgements start here
%----------------------------------------------
We thank CERN for the very successful operation of the LHC, as well as the
support staff from our institutions without whom ATLAS could not be
operated efficiently.

We acknowledge the support of ANPCyT, Argentina; YerPhI, Armenia; ARC, Australia; BMWFW and FWF, Austria; ANAS, Azerbaijan; SSTC, Belarus; CNPq and FAPESP, Brazil; NSERC, NRC and CFI, Canada; CERN; CONICYT, Chile; CAS, MOST and NSFC, China; COLCIENCIAS, Colombia; MSMT CR, MPO CR and VSC CR, Czech Republic; DNRF and DNSRC, Denmark; IN2P3-CNRS, CEA-DSM/IRFU, France; SRNSF, Georgia; BMBF, HGF, and MPG, Germany; GSRT, Greece; RGC, Hong Kong SAR, China; ISF, I-CORE and Benoziyo Center, Israel; INFN, Italy; MEXT and JSPS, Japan; CNRST, Morocco; NWO, Netherlands; RCN, Norway; MNiSW and NCN, Poland; FCT, Portugal; MNE/IFA, Romania; MES of Russia and NRC KI, Russian Federation; JINR; MESTD, Serbia; MSSR, Slovakia; ARRS and MIZ\v{S}, Slovenia; DST/NRF, South Africa; MINECO, Spain; SRC and Wallenberg Foundation, Sweden; SERI, SNSF and Cantons of Bern and Geneva, Switzerland; MOST, Taiwan; TAEK, Turkey; STFC, United Kingdom; DOE and NSF, United States of America. In addition, individual groups and members have received support from BCKDF, the Canada Council, CANARIE, CRC, Compute Canada, FQRNT, and the Ontario Innovation Trust, Canada; EPLANET, ERC, ERDF, FP7, Horizon 2020 and Marie Sk{\l}odowska-Curie Actions, European Union; Investissements d'Avenir Labex and Idex, ANR, R{\'e}gion Auvergne and Fondation Partager le Savoir, France; DFG and AvH Foundation, Germany; Herakleitos, Thales and Aristeia programmes co-financed by EU-ESF and the Greek NSRF; BSF, GIF and Minerva, Israel; BRF, Norway; CERCA Programme Generalitat de Catalunya, Generalitat Valenciana, Spain; the Royal Society and Leverhulme Trust, United Kingdom.

The crucial computing support from all WLCG partners is acknowledged gratefully, in particular from CERN, the ATLAS Tier-1 facilities at TRIUMF (Canada), NDGF (Denmark, Norway, Sweden), CC-IN2P3 (France), KIT/GridKA (Germany), INFN-CNAF (Italy), NL-T1 (Netherlands), PIC (Spain), ASGC (Taiwan), RAL (UK) and BNL (USA), the Tier-2 facilities worldwide and large non-WLCG resource providers. Major contributors of computing resources are listed in Ref.~\cite{ATL-GEN-PUB-2016-002}.
%----------------------------------------------

\printbibliography

\newpage
% ATLAS Collaboration author list
% Data extracted on 23-Jan-2017 for paper reference STDM-2016-08

\begin{flushleft}
{\Large The ATLAS Collaboration}

\bigskip

M.~Aaboud$^\textrm{\scriptsize 137d}$,
G.~Aad$^\textrm{\scriptsize 88}$,
B.~Abbott$^\textrm{\scriptsize 115}$,
J.~Abdallah$^\textrm{\scriptsize 8}$,
O.~Abdinov$^\textrm{\scriptsize 12}$,
B.~Abeloos$^\textrm{\scriptsize 119}$,
S.H.~Abidi$^\textrm{\scriptsize 161}$,
O.S.~AbouZeid$^\textrm{\scriptsize 139}$,
N.L.~Abraham$^\textrm{\scriptsize 151}$,
H.~Abramowicz$^\textrm{\scriptsize 155}$
% ...
\bigskip
\\
% ...
$^{8}$ Department of Physics, The University of Texas at Arlington, Arlington TX, United States of America\\
% ...
$^{12}$ Institute of Physics, Azerbaijan Academy of Sciences, Baku, Azerbaijan\\
% ...
$^{88}$ CPPM, Aix-Marseille Universit{\'e} and CNRS/IN2P3, Marseille, France\\
% ...
$^{115}$ Homer L. Dodge Department of Physics and Astronomy, University of Oklahoma, Norman OK, United States of America\\
% ...
$^{119}$ LAL, Univ. Paris-Sud, CNRS/IN2P3, Universit{\'e} Paris-Saclay, Orsay, France\\
% ...
$^{137}$ $^{(a)}$ Facult{\'e} des Sciences Ain Chock, R{\'e}seau Universitaire de Physique des Hautes Energies - Universit{\'e} Hassan II, Casablanca; $^{(b)}$ Centre National de l'Energie des Sciences Techniques Nucleaires, Rabat; $^{(c)}$ Facult{\'e} des Sciences Semlalia, Universit{\'e} Cadi Ayyad, LPHEA-Marrakech; $^{(d)}$ Facult{\'e} des Sciences, Universit{\'e} Mohamed Premier and LPTPM, Oujda; $^{(e)}$ Facult{\'e} des sciences, Universit{\'e} Mohammed V, Rabat, Morocco\\
% ...
$^{139}$ Santa Cruz Institute for Particle Physics, University of California Santa Cruz, Santa Cruz CA, United States of America\\
% ...
$^{151}$ Department of Physics and Astronomy, University of Sussex, Brighton, United Kingdom\\
% ...
$^{155}$ Raymond and Beverly Sackler School of Physics and Astronomy, Tel Aviv University, Tel Aviv, Israel\\
% ...
$^{161}$ Department of Physics, University of Toronto, Toronto ON, Canada\\
% ...
$^{a}$ Also at Department of Physics, King's College London, London, United Kingdom\\
$^{b}$ Also at Institute of Physics, Azerbaijan Academy of Sciences, Baku, Azerbaijan\\
$^{c}$ Also at Novosibirsk State University, Novosibirsk, Russia\\
$^{d}$ Also at TRIUMF, Vancouver BC, Canada\\
$^{e}$ Also at Department of Physics {\&} Astronomy, University of Louisville, Louisville, KY, United States of America\\
% ...
\end{flushleft}

% Created with xml2latex.py

\end{document}